\documentclass[nonacm,sigconf]{acmart}

\AtBeginDocument{%
  \providecommand\BibTeX{{%
    \normalfont B\kern-0.5em{\scshape i\kern-0.25em b}\kern-0.8em\TeX}}}
\copyrightyear{2022}
\acmYear{2022}
\acmConference[WSDM  '22 (submitted)]{WSDM '22: ACM International Conference on Web Search and Data Mining}{February 21--25, 2022}{Phoenix, AZ}
\acmBooktitle{WSDM  '22 (submitted): ACM International Conference on Web Search and Data Mining}
\usepackage{times}
\usepackage{latexsym}
\usepackage{xcolor}
\usepackage{xspace}

\usepackage{booktabs,tabularx, colortbl}
\usepackage{microtype}
\usepackage{graphicx}
\usepackage{multirow}
\usepackage{amsmath}
\usepackage{amsfonts}
\usepackage{amsthm}
\usepackage{bm}
\usepackage{cancel}
\usepackage{algorithm,algorithmicx,algpseudocode}
\usepackage{multicol}
\usepackage{enumitem}
\usepackage{thm-restate}
\usepackage{makecell}
\usepackage{units}
\usepackage{xcolor,colortbl}
\usepackage{adjustbox}
\raggedcolumns

\newtheorem{definition}{Definition}
\newcommand\BibTeX{B\textsc{ib}\TeX}

\newif\ifcorecom

\corecomtrue  

\ifcorecom
	\newcommand{\core}[1]{{#1}}
\else
    \newcommand{\core}[1]{}
\fi

\newcommand{\nachshon}[1]{\core{\color{blue}{{\bf Nachshon: }#1}}}
\newcommand{\amit}[1]{\core{\color{violet}{{\bf Amit: }#1}}}

\newcommand{\alex}[1]{{\color{brown}{{\bf Alex: }#1}}}
\newcommand{\oren}[1]{{\color{cyan}{{\bf Oren: }#1}}}

\renewcommand{\alex}[1]{}
\renewcommand{\oren}[1]{}

\newcommand{\ignore}[1]{}
\newcommand{\shorten}[1]{}
\newcommand{\T}[1]{\noindent\textbf{#1}}

\newcommand{\Ours}{AutoEncoder with Side Information\xspace}

\newcommand{\oursabbr}{\mbox{AESI}\xspace}
\newcommand{\splitbert}{BERT\textsubscript{\textsc{SPLIT}}\xspace}
\newcommand{\norm}[1]{\left\lVert#1\right\rVert}

\DeclareMathOperator*{\argmin}{arg\,min}

\title{SDR: Efficient Neural Re-ranking using Succinct Document Representation}

\author{Nachshon Cohen\footnotemark[1], Amit Portnoy\footnotemark[1], Besnik Fetahu, and Amir Ingber}

\date{}

\begin{document}

\begin{abstract}

BERT based ranking models have achieved superior performance on various information retrieval tasks. However, the large number of parameters and complex self-attention operation come at a significant latency overhead. To remedy this, recent works propose late-interaction architectures, which allow pre-computation of intermediate document representations, thus reducing the runtime latency. Nonetheless, having solved the immediate latency issue, these methods now introduce storage costs and network fetching latency, which limits their adoption in real-life production systems.

In this work, we propose the Succinct Document Representation (SDR) scheme that computes \emph{highly compressed} intermediate document representations, mitigating the storage/network issue. Our approach first reduces the dimension of token representations by encoding them using a novel autoencoder architecture that uses the document's textual content in both the encoding and decoding phases. After this token encoding step, we further reduce the size of entire document representations using a modern quantization technique. 

Extensive evaluations on passage re-reranking on the MSMARCO dataset show that compared to existing approaches using compressed document representations, our method is highly efficient, achieving 4x--11.6x better compression rates for the same ranking quality.

\end{abstract}
\maketitle
\renewcommand{\shortauthors}{Cohen et al.}
\renewcommand{\thefootnote}{\fnsymbol{footnote}}
\footnotetext[1]{Both authors contributed equally to the paper.}
\renewcommand{\thefootnote}{\arabic{footnote}}

\section{Introduction}

Information retrieval (IR) systems traditionally comprise of two stages: retrieval and ranking. Given a user query, the role of the retrieval stage is to quickly retrieve a set of candidate documents from a (very large) search index. The retrieval algorithm is typically fast but not accurate enough; in order to improve the quality of the end result for the user, the candidate documents are re-ranked using a more accurate but computationally expensive algorithm.

Large deep learning models have achieved the state of the art ranking performance in IR applications~\cite{yates-etal-2021-pretrained}. Transformer networks such as BERT \cite{devlinetal_bert} consistently show better ranking effectiveness at the cost of a higher computational cost and latency \cite{Nogueira_BERT_passage}.

% In some applications, the latency requirements are so strict that using BERT for ranking is not possible at all.

%Re-ranking is a crucial part of information retrieval systems. Given a subset of documents returned by the retriever, it is possible to apply advanced ML models to decides on the final rank of the documents presented to the user. 
%The re-ranking quality has significant impact on the overall quality of the system, so it is generally important to apply high-quality re-ranking models, such as Bert. 
%However, latency is also an important factor of a retrieval system, which limits the type of models that can be used for re-ranking. Specifically, running a Bert model on all documents returned by the retrieval can be prohibitively expensive. 

\begin{figure}
    \centering
    \includegraphics[width=1.1\columnwidth]{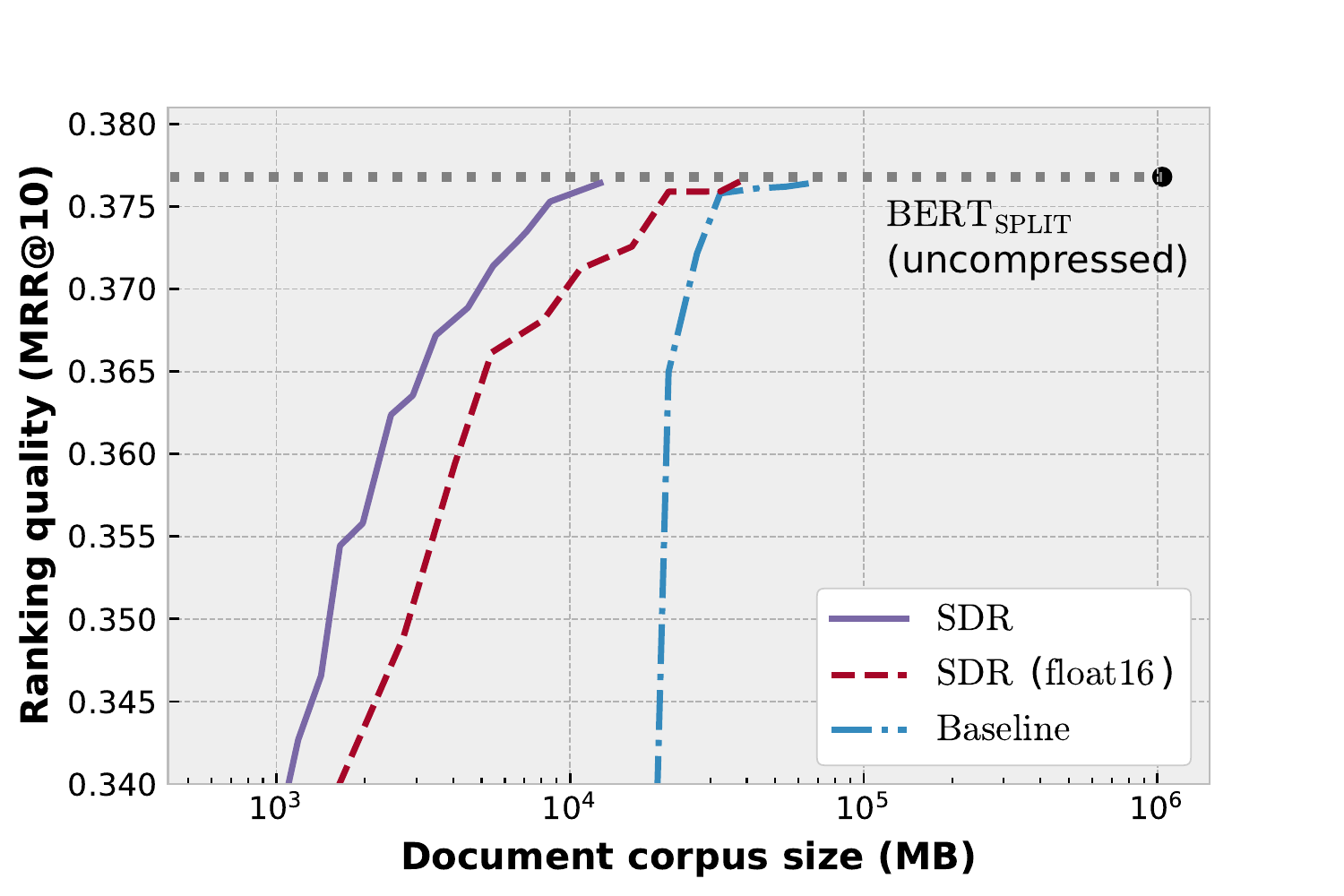}
    \caption{\emph{High-level contribution}: MRR@10 performance vs. document corpus size tradeoff, measured on the MSMARCO-DEV dataset. \splitbert is a distilled late-interaction model with reduced vector width and no compression (\S~\ref{subsection:baseline}).
%   The `State-of-the-Art' line represents a two layer autoencoder (\S~\ref{subsec:auto-encoder-mrr}, \textbf{AE-2L}) and the two SDR lines use our novel autoencoder (\S~~\ref{subsec:autoencoder}, \textbf{AESI}). For quantization, `SDR' also uses an advance quantization method we describe in \S~\ref{subsec:quantization-algorithm}, while the other two use float16 quantization, which is a natural improvement upon the suggestion in \cite{macavaney2020efficient}. %note that macavaney2020efficient used 1layer AE on BERT-base. We use a much stronger baseline with 2layer AE on distilled BERT. 
%   The lines are are derived from efficient configurations of varying encoding widths. 
    For MRR@10 above 0.35, SDR is 4x--11.6x more efficient compared to the baseline. 
    % Our method, SDR, is \textasciitilde10x more efficient when using float16 quantization and, when using a more advanced quantization scheme, as we describe in \ref{subsec:quantization-algorithm}, achieves an improvement of \textasciitilde20x.\amit{todo refer here}
    %   \WIP{The state-of-the-art line represents a two layer autoencoder with varying encoding widths (\S~\ref{subsec:auto-encoder-mrr}) and float16 quantization, a natural improvement over \cite{macavaney2020efficient}.
    % \amit{<- after the previous edit, we are not explaining SDR (float16), was it intentional?} }
    }
    \label{fig:contrib-overview}
\end{figure}

\begin{figure}
    \centering
    \includegraphics[width=1.1\columnwidth]{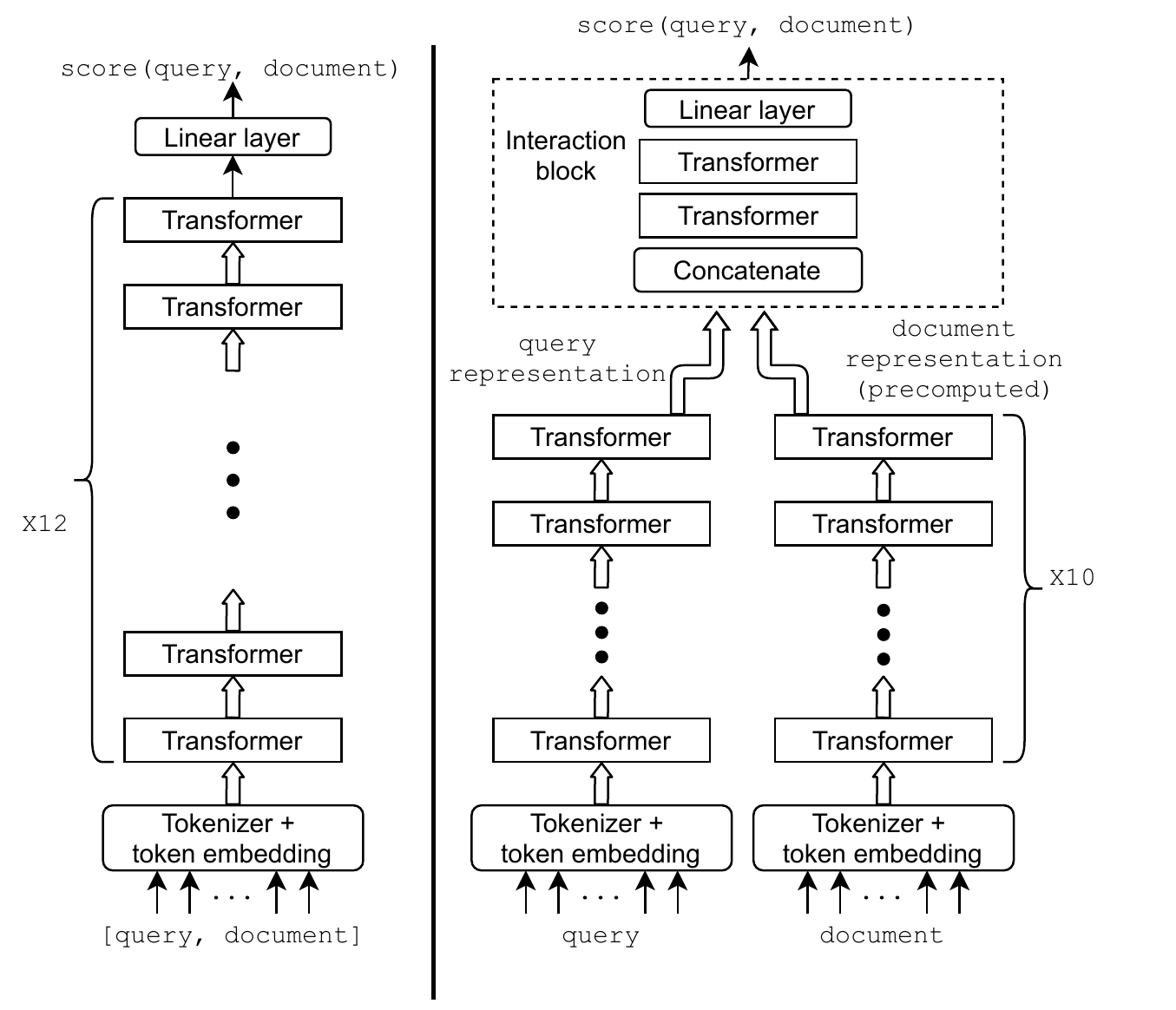}
    \caption{Left: BERT ranker. Right: late-interaction ranker (with two transformer layers as the interaction block). \ignore{\oren{perhaps add and describe the entire flow. Took me a while to understand why you're mentioning the ES index size, ...}}}
    \label{fig:late-interaction}
\end{figure}

% \vspace{-0.05cm}
To rank $k$ documents, the ranker is called $k$ times with an input of the form (query, document), where the query is the same, but the document is different. Several works~\cite{macavaney2020efficient,gao2020modularized,chen2020dipair,cao2020deformer,nie2020dc,gao2020modularized,khattab2020colbert} have proposed to modify BERT-based rankers in a way that allows part of the model to compute query and document representations separately, and then produce the final score using a low-complexity interaction block (we denote these models as late-interaction rankers, see Fig.~\ref{fig:late-interaction}). With this approach, the document representations can be pre-computed in order to improve latency significantly: during runtime, the model computes the query representation (once), retrieves the pre-computed document representations, and only runs the interaction block $k$ times to produce the final ranking score.

Precomputing document representations have been shown to significantly reduce latency while at the same time retaining comparable scores to BERT models~\cite{gao2020modularized}. However, this does not account for additional storage and/or network fetching latency costs: these representations typically consist of the contextual token embeddings in a transformer model, which translate to orders of magnitude larger space requirements than storing the entire corpus search index\footnote{In Appendix~\ref{subsec:analysis-elasticsearch}, we also demonstrate how the latency introduced by such an increase in storage requirements can dominate the end-to-end latency.}.

In this work, we propose Succinct Document Representation (SDR), a general scheme for compressing document representations. Our scheme enables late-interaction rankers to be efficient in both latency and storage while maintaining high ranking quality. SDR is suitable for any ranking scheme that relies on contextual embeddings and achieves extreme compression ratios (2-3 orders of magnitude) with little to no impact on retrieval accuracy. SDR consists of two major components: (1) embedding dimension reduction using an \emph{autoencoder} with \emph{side information} and (2) \emph{distribution-optimized quantization} of the reduced-dimension vectors.

In SDR, the autoencoder consists of two subnetworks: an encoder that reduces the vector's dimensions and a decoder that reconstructs the compressed vector. The encoder's output dimension represents the tradeoff between reconstruction fidelity and storage requirements. To improve the compression-reliability tradeoff, we leverage \emph{static} token embeddings, 
%\oren{What are those? You mean the "static word embeddings"? There are works on combining static and contextualized word vectors which shows that they improve results (even a library called Flair). Still legit, but should reference.} 
which are available given that the ranker has access to the document text (as it needs to render it to the user), and are computationally cheap to obtain. We feed these embeddings to both the encoder and decoder as \emph{side information},
%\oren{No longer an autoencoder, since input!=output, right? a context encoder perhaps...}
allowing the autoencoder to focus more on storing \emph{``just the context''} of a token, and less on its original meaning that is available in the \emph{static} embeddings. Ablation tests verify that adding the static vectors significantly improves the compression rates for the same ranker accuracy.

% An autoencoder is a neural network that consists of two subnetworks - an encoder, which takes an input vector and outputs a reduced dimension vector, and a decoder which attempts to reconstruct the original input vector. The target dimension of the encoder output controls the tradeoff between reconstruction fidelity and storage requirements. In SDR, the autoencoder takes advantage of the fact that the vectors to be compressed are contextual embeddings in a transformers model, which generally represent the semantics of the word in a high-dimensional space within the context of the other words in the document. Since the ranker always has access to the document text (it needs to be presented to the user), it has access to the original -- uncontextualized -- word embedding of each token. We therefore feed the uncontextualized vectors to both the encoder and the decoder as side information, so the reduced-dimension vector can focus on storing ``just the context'' of a word, without its original meaning. Ablation tests verify that adding the uncontextualized vectors significantly improves the compression rates for the same ranker accuracy.

Since data storage is measured in bits rather than floating-point numbers, SDR uses quantization techniques to reduce storage size further. Given that it is hard to evaluate the amount of information in each of the encoder's output dimensions, we perform a randomized Hadamard transform on the vectors, resulting in (1) evenly spread information across all coordinates and (2) transformed vectors that follow a Gaussian-like distribution. We utilize known quantization techniques to represent these vectors using a small number of bits, controlling for the amount of quantization distortion.

Existing late-interaction schemes either ignore the storage overhead, or consider basic compression techniques, such as a simple (1 layer) autoencoder and float16 quantization.
However, this is insufficient to reach reasonable storage size \cite{macavaney2020efficient}; furthermore, fetching latency is unacceptable for interactive systems (cf. Appendix~\ref{subsec:analysis-elasticsearch}). 
As an underlying late-interaction model we use a distilled model with a reduced vector width \cite{hofstatter2020improving}. 
As a baseline compression scheme, we use a non-linear autoencoder consisting of 2 dense layers followed by float16 quantization, a natural extension of \cite{macavaney2020efficient}.
On the MSMARCO dataset, this baseline achieves compression rates of 30x with no noticeable reduction in retrieval accuracy (measured with the official MRR@10 metric). 
% We therefore consider as a baseline a distilled BERT model with a reduced vector width, and a non-linear autoencoder consisting of 2 dense layers. 
% On the MSMARCO dataset, this baseline achieves compression rates of 30x compared to the BERT vectors with no noticeable reduction in retrieval accuracy (measured with the official MRR@10 metric). 
On top of this strong baseline, our SDR scheme achieves an additional compression rate of between 4x to 11.6x with the same ranking quality, reducing document representation size to the same order of magnitude as the retrieved text itself. In Figure~\ref{fig:contrib-overview} we include a high-level presentation of the baseline, a variant of our method with float16 quantization, and our full method.

To summarize, here are the contribution of this work:
\begin{itemize}[leftmargin=*]
    \item We propose the Succinct Document Representation (SDR) scheme for compressing the document representations required for fast Transformer-based rankers. The scheme is based on a specialized autoencoder architecture and subsequent quantization.
    \item For the MSMARCO passage retrieval task, SDR shows compression ratios of 121x with no noticeable decrease in ranking performance. Compared to existing approaches for producing compressed representations, our method attains better compression rates (between 4x and 11.6x) for the same ranking quality.
    %\amir{add results on any other datasets, if we have them}\amit{@besnik add nDCG results on MSMARCO-TREC19}
    \item We provide a thorough analysis of the SDR system, showing that the contribution of each of the components to the compression-ranking effectiveness is significant.
\end{itemize}

\section{Related Work}

\textbf{Late-interaction models.} The idea of running several transformer layers for the document and the query independently, and then combining them in the last transformer layers, was developed concurrently by multiple teams: PreTTR \cite{macavaney2020efficient}, EARL \cite{gao2020earl}, DC-BERT \cite{nie2020dc}, DiPair \cite{chen2020dipair}, and the Deformer \cite{cao2020deformer}.
% We denote this type of architecture as \splitbert. \amit{@nacshon, we later talk about \splitbert as though it is a single clear arch that we can sample results from, here it's described as a type of arch and it's not clear exactly what \splitbert is.}
% \amir{why not simply call these "late interaction models" here, and when we want to refer to a specific one, state which one?}
These works show that only a few layers where the query and document interact are sufficient to achieve results close to the performance of a full BERT ranker at a fraction of the runtime cost. For each document, the contextual token vectors are stored in a cache and retrieved during the document ranking phase.  This impacts both storage cost (storing token contextual vectors for all documents) as well as latency cost of fetching these vectors during the ranking phase.
MORES \cite{gao2020modularized} is an extension of the late-interaction models, where in the last interaction layers, the query attends to the document, but not vice versa, and without document self-attention. As the document is typically much longer, this modification results in an additional performance improvement with similar storage requirements. ColBERT~\cite{khattab2020colbert} is another variant that runs all transformer layers independently for the query and the document, and the interaction between the final vectors is done through a sum-of-max operator.
In a similar line of work, the Transformer-Kernel (TK)~\cite{hofstatter2020interpretable}, has an interaction block based on a low-complexity kernel operation. 
%\amit{<- @amir, can we add something beyond "kernels" (the current phrasing open too many questions)} 
Both ColBERT and TK result in models with lower runtime latency at the expense of a drop in ranking quality. However, the storage requirements for both approaches are still significant.

% ColBERT~\cite{khattab2020colbert} is another variant that runs all transformer layers independently for the query and the document, and the interaction between the final vectors is done through a sum-of-max operator
% \amir{move this to the related work} \WIP{Some researchers suggested minimizing such latency induced by large transformer networks by designing a more efficient ranking schemes at the expense of ranking quality~\cite{Hofstatter_runtime}. Given the same time allotted for computation per query, a faster ranker can operate on more documents than a less efficient one, leading to higher overall effectiveness \cite{hofstatter2020interpretable}.}

Some of the works above acknowledge the issue of storing the precomputed document representations and proposed partial solutions. In ColBERT~\cite{khattab2020colbert}, the authors proposed to reduce the dimension of the final token embedding using a linear layer. However, even moderate compression ratios caused a drop in ranking quality. 
%offers the possibility to reduce storage requirements by reducing the vectors' dimensions, but again, drastic reduction causes a significant drop in quality. 
In the PreTTR model~\cite{macavaney2020efficient}, it was proposed to address the storage cost by using a standard auto-encoder architecture and the float16 format instead of float32. Again, the ranking quality drops even with moderate compression ratios (they measured up to 12x).

%However, they only reached compression rates of up to 12x, which is still insufficient as the vectors even with 12x compression consume large amounts of storage and impact retrieval latency.  

Several other works \cite{guu2020retrieval,karpukhin2020dense,xiong2021approximate,qu2020rocketqa,lu2020twinbert} proposed representing the queries and documents as vectors (as opposed to a vector per token), and using a simple function (e.g., the dot product or cosine distance) as the interaction block. While this ranker architecture approach is simple (and can also be used for the retrieval step via an approximate nearest neighbor search such as FAISS~\cite{johnson2017billion} or ScaNN \cite{avq_2020}), the overall ranking quality is generally lower compared to methods that employ a query-document cross-attention interaction.

%Multiple other works have proposed to 

%REALM \cite{guu2020retrieval}, DPR \cite{karpukhin2020dense}, ANCE \cite{xiong2021approximate}, RocketQA \cite{qu2020rocketqa}, and TwinBERT \cite{lu2020twinbert} all encode the query and document as a single vector. 

%The query-document interaction is done via a simple dot product. Such methods can be used also for the retrieval step via an approximate nearest neighbor search such as FAISS~\cite{johnson2017billion} or ScaNN \cite{avq_2020}. 
%However, when used without an additional ranking step, the quality is lower compared to methods that employ a query-document cross-attention interaction. 

\textbf{Knowledge distillation.} Knowledge distillation techniques, such as DistilBERT \cite{sanh2019distilbert} and TinyBERT \cite{jiao2020tinybert}, can reduce the size of a model at a small cost in terms of quality.
Such works were successfully applied for ranking~\cite{hofstatter2020improving,chen2021simplified} and do not require storing document vectors. However, distillation works best when the distilled model has at least 6 layers. With just 2-3 layers, late interaction models generally achieve better quality than distillation models by smartly using the precomputed document representations.

\textbf{Compressed embeddings.} Our work reduces storage requirements by reducing the number of bits per floating-point value. Quantization gained attention and success in reducing the size of neural network parameters \cite{pmlr-v37-gupta15, 8008430, wang2018training, wu2018training} and distributed learning communication costs \cite{pmlr-v70-suresh17a, NIPS2017_6c340f25, konevcny2018randomized, vargaftik2021drive}. Specifically, compressing word embeddings has been studied as an independent goal. May et al.~\cite{NIPS2019_compressed_embeddings} studied the effect of quantized word embeddings on downstream applications and proposed a metric for quantifying this effect with simple linear models that operate on the word embeddings directly. As our work is concerned with compressing \emph{contextual} embeddings, these methods do not apply since the set of possible embeddings values is not bounded by the vocabulary size. Nevertheless, as in \cite{NIPS2019_compressed_embeddings}, we also observe that simple quantization schemes are quite effective. Our work uses recent advances in this area to further reduce storage requirements for document representation, which, to the best of our knowledge, were not previously attempted in this context. 

\section{Succinct Document Representation (SDR)}\label{sec:algorithm}
Our work is based on the late-interaction architecture \cite{macavaney2020efficient,gao2020modularized,chen2020dipair,cao2020deformer,nie2020dc}, which separates BERT into $L$ independent layers for the documents and the queries, and $T-L$ interleaving layers, where $T$ is the total number of layers in the original model, e.g., 12 for BERT-Base. Naively storing all documents embeddings consumes a huge amount of storage with a total of  $m\cdot h \cdot 4$ bytes per document, where $m$ is the average number of tokens per document and $h$ is the model hidden size (384 for the distilled version we use). For MSMARCO, with 8.8M documents and $m\!=\!76.9$, it leads to a high storage cost of over a terabyte, which is not affordable except in large production systems.

Our compression scheme for the document representations consists of two sequential steps, (i) dimensionality reduction and (ii) block-wise quantization, described in \S~\ref{subsec:autoencoder} and \S~\ref{subsec:quantization-algorithm} respectively. \ignore{In \S~\ref{subsec:overall-algo} we describe the overall system and parameter choices.} 
%Later, in \S~\ref{sec:eval}, we evaluate SDR and its components and compare with alternative methods.

%\alex{I'm missing a figure that shows overall arch flow \nachshon{Would be good to add depending on time constraint}} 

\subsection{Dimensionality Reduction using an AutoEncoder with Side Information (AESI)}\label{subsec:autoencoder}

To compress document representations, we reduce the dimensionality of token representations (i.e., the output of BERT's $L$-th layer) using an autoencoder. 
Standard autoencoder architectures typically consist of a neural network split into an encoder and a decoder: the encoder projects the input vector into a lower-dimension vector, which is then reconstructed back using the decoder. 
%. For an input vector (e.g., BERT's $L$-th output layer), the encoder projects a lower-dimension vector, which is reconstructed back during the decoding phase. 

\ignore{
\amit{the following paragraph seems to not add much new information, wrapping in \textbackslash WIP for now}
There is an inherent trade-off between compression rate and the quality of the reconstructed vectors. Hence, we formalize two desiderata that our compression algorithm needs to fulfill: \amit{seems a bit redundant to say both desiderata and 'needs to fulfill'}
\begin{enumerate}[leftmargin=*]
\item High document representation compression rate that significantly reduces storage and latency costs.
\item High reconstruction fidelity of compressed representation without harming the utility in downstream tasks.
\end{enumerate}
}

% While autoencoders can achieve good dimensionality reduction, it reduces quality drastically when the projected size is too small. In such a case, the autoencoder has to encode all the data in the original vector in just a few dimensions of the projected vector, reducing reconstruction accuracy. 

%%%% THIS IS WHERE I STOPPED. I THINK THE SECTION NEEDS TO BE REORGANIZED. CURRENTLY IT CONTAINS INFORMATION FROM EXP. SETUP AND ALSO THE ARGUMENTATION IS A BIT UNSTRUCTURED.
%Autoencoders are trained by minimizing the mean squared error (MSE) between the reconstructed vector and the original vector. 

Our architecture, AESI, extends the standard autoencoder by using the document's text as \emph{side information} to both the encoder and decoder. Such an approach is possible since, no matter how the document scores are computed, re-ranking systems have access to the document's text in order to render it back to the user. In the rest of this section, we add the precise details of the AESI architecture.

\ignore{
\amit{the following paragraph seems to not add much new information, wrapping in \textbackslash WIP for now}
The document's textual content can be utilized in two main ways: (i) precompute representations and cache them, thus, inducing significant storage costs, and (ii) compute the representations from raw text by running it through all $L$ transformer layers, thus, inducing high latency costs. Our approach, \oursabbr{}, finds a middle ground between \emph{caching} an little amount of pre-computed document information, and performing a lightweight \emph{online computation}.
}

% Our key observation is that ranking system must have access to the document text, which is eventually returned to the user. The text encodes useful information, e.g., by running the text through $L$ BERT layers, it is possible to fully reconstruct the original embeddings without storing any data.  Of course, running $L$ BERT layers during online inference is time consuming, and defeats the entire purpose of speeding up the ranker.  Given document embedding vectors and raw text, we can consider two extremes: (1) Discard the raw text and cache the document embeddings; and (2) Discard the document embedding and compute it from the raw text.  The former is heavy in storage, while the latter is heavy in computation.  Our aim is to find a middle ground, where we cache a small amount of information and perform some lightweight computation. 

\paragraph{\textbf{Side Information.}} In line with our observation that we have access to the document's raw text, we propose utilizing the \emph{token embedding} information, which is computed by the embedding layer used in BERT's architecture. The token embeddings encode rich semantic information about the token itself; however, they do not fully capture the context in which they occur; hence, we refer to them as \emph{static embeddings}. For example, through token embeddings, we cannot disambiguate between the different meanings of the token \texttt{bank}, which can refer to either a geographical location (e.g., ``river bank'') or a financial institution, depending on the context.

%\emph{``bank''} \guy{I think tokens are usually written without "}, which depending on the context, can refer to \emph{geographical location} (e.g., ``river bank'') or \emph{financial institution}. 

% \paragraph{\textbf{Side Information.}} Inline with our observation that we have access to document's raw text, we propose in utilizing the \emph{token embedding} information, which the first input layer (Layer 0) used in BERT's multi-layer transformer architecture. Token embeddings encode rich semantic information about the word itself, however, as such they do not fully capture the context in which they occur, hence we refer to them as \emph{uncontextual embeddings}. For example, through token embeddings we cannot disambiguate the correct \emph{word sense} of the token \emph{``bank''}, which depending on the context, can refer to \emph{geographical location} (e.g., ``river bank'') or \emph{financial institution}. 

Static embeddings are key for upper BERT layers, which learn the contextual representation of tokens via the self-attention mechanism.%, by taking into account all the possible interactions that a token can have with other tokens in a given text sequence. 
We use the static embeddings as side information to both the encoder and decoder parts of the autoencoder. This allows the model to focus on encoding the \emph{distilled context}, and less on the token information since it is already provided to the decoder directly. Encoding only the context is an easier task, allowing AESI to achieve higher compression rates, and at the same time, retaining the quality of the contextual representation.

\begin{figure}
    \centering
    \includegraphics[width=.8\columnwidth]{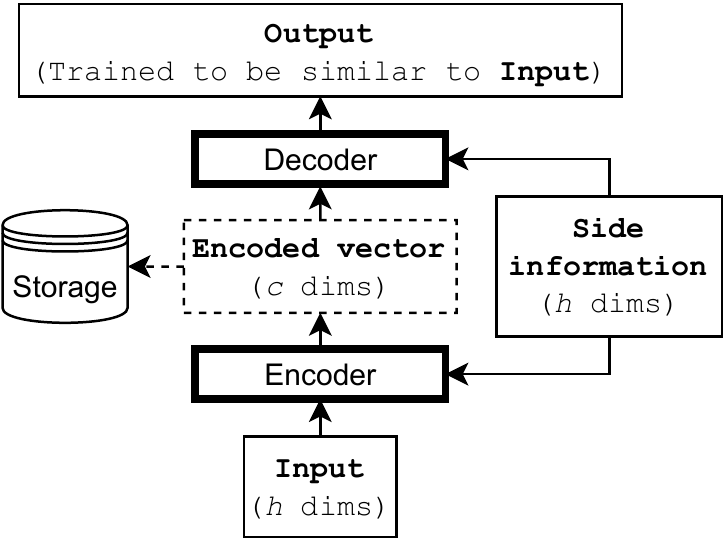}
    \caption{\Ours{} (\oursabbr) architecture. For our usage, the input is the contextual token embedding (the \mbox{$L$-th} layer's output), and the side information is the static token embedding (the output of BERT's initial embedding layer). The resulting $c$--dimensional encoded vector can be thought of as the distilled context of the input token.
    % \amir{1. let's merge the 'concat' into the encoder (i.e. no separate block). 2. the side information should enter the encoder from the side :-) 3. how about 'token embedding' instead of layer 0? }
    }
    \label{fig:autoencoder-arch-with-uncontext}
\end{figure}

\paragraph{\textbf{AESI Approach.}} For a token whose representation we wish to compress, our approach proceeds as follows. We take the $L$-th layer's output contextual representation of the token together with its static embedding and feed both inputs to the autoencoder. The information to be compressed (and reconstructed) is the contextual embedding, and the side-information, which aids in the compression task, is the static embedding. The decoder takes the encoder output, along with the static embedding, and attempts to reconstruct the contextual embedding. Figure~\ref{fig:autoencoder-arch-with-uncontext} shows the AESI architecture.

%The encoder's output, which represents the compressed contextual representation, during the decoding phase is reconstructed back by additionally utilizing the uncontextual embeddings, allowing thus, to have higher reconstruction fidelity. 

With the AESI approach, there are several parameters that we determine empirically. First, the $L$-th transformer layer of the contextual representation that is provided as input, which has a direct impact on latency\footnote{A ranker model has to compute layers $L+1$ onward online.}. Second, the size of the encoder's output directly impacts the compression rate and thus storage costs.

%Figure~\ref{fig:autoencoder-arch-with-uncontext} shows the architecture of our AESI compression approach. 
Encoding starts by concatenating
%\guy{probably to late if you haven't tried this, but isn't it more intuitive to subtract them? the L-vector contains info on both token and context, the static contains only token info and we want only the context info \amit{if this is the relation between the layers the autoencoder would have probably "encode" it anyway no?}} 
the input vector (i.e., the output of layer $L$, the vector we compress) and the static token embedding (i.e., the output of BERT's embedding layer), and then passes the concatenated vector through an encoder network, which outputs a $c$-dimensional {\em encoded vector}. 
Decoding starts by concatenating the encoded vector with the static token embedding, then passes the concatenated vector through a decoder layer, which reconstructs the input vector.
Specifically, we use a two-layer dense network for both the encoder and the decoder, which can be written using the following formula:
\begin{align}
e=E(v, u) := W^e_2\cdot\bigl(\operatorname{gelu}\bigl(W^e_1(v;u)\bigr)\bigr) 
\\
v'=D(e, u) := W^d_2\cdot\bigl(\operatorname{gelu}\bigl(W^d_1(e;u)\bigr)\bigr)
\end{align}
where $v\in\mathbb{R}^h$ is the contextualized token embedding (the output of the $L$-th layer), $u\in\mathbb{R}^h$ is the static token embedding (the output of the embedding layer, which is the input to BERT's layer $0$ and includes token position embeddings and type embeddings), and $u;v$ means concatenation of these vectors. \mbox{$W^e_1\in\mathbb{R}^{i\times 2h}$}, \mbox{$W^e_2\in\mathbb{R}^{c\times i}$}, \mbox{$W^d_1\in\mathbb{R}^{i\times (c+h)}$}, \mbox{$W^d_2 \in\mathbb{R}^{h\times i}$} are trainable parameters. 
$h$ is the dimension of token embeddings (e.g., 384), $i$ is the intermediate autoencoder size, and $c$ is the dimension of the projected (encoded) vector. $\operatorname{gelu}(\cdot)$ is an non-linear activation function \cite{hendrycks2016gaussian}. Additional autoencoder variations are explored in \S~\ref{subsec:auto-encoder-mrr}. 

\subsection{Quantization}\label{subsec:quantization-algorithm}

Storing the compressed contextual representations in a naive way consumes 32 bits (float32) per coordinate per token, which is still costly. To further reduce storage overhead, we propose to apply a quantization technique, which uses a predetermined $B$ bits per coordinate. We rely on a recently proposed quantization approach, DRIVE~\cite{vargaftik2021drive}, which we describe next and summarize in Algorithm~\ref{code:ourQ}. Later in this subsection, we show how we apply DRIVE to our AESI-encoded documents.

% We start by presenting the quantization scheme on a fixed-length vector, which is based on a recent algorithm, nicknamed DRIVE~\cite{vargaftik2021drive}, and is summarized in Algorithm \ref{code:ourQ}.  Later in this section, we will describe how to apply this method to a batch of (auto)encoded variable-length input vectors.

% Once the input vector is encoded (via the encoder), it is possible to store it directly in storage. However, naively storing the vectors in float32 format consumes 32 bits per coordinate per token, which is still expensive. To further reduce storage overhead, we quantize the vectors into $B$ bits per coordinate while preserving most of the important information.

% We start by presenting the quantization scheme on a fixed-length vector, which is based on a recent algorithm, nicknamed DRIVE~\cite{vargaftik2021drive}, and is summarized in Algorithm \ref{code:ourQ}. 
% Later in this section, we will describe how to apply this method to a batch of (auto)encoded variable-length input vectors.
% \nachshon{Add that the length is variable? \amit{added}}

% \noindent \textbf{Hadamard preconditioning.} 
Before going into the details of the quantization method, we first require the following definitions:

\begin{definition}[\cite{horadam2012hadamard}]
 A normalized Walsh-Hadamard matrix,\\ \mbox{$H_{2^k} \in \{+1,-1\}^{2^k\times2^k}$}, is recursively defined as
\begin{gather*}
H_1 = 1;\ \ H_{2^k} = \frac{1}{\sqrt{2}}\begin{pmatrix}
H_{2^{k-1}} & H_{2^{k-1}} \\
H_{2^{k-1}} & -H_{2^{k-1}}
\end{pmatrix} .
\end{gather*}
% \mbox{ 
% $ \small
% H_{2^k}{=} \begin{pmatrix}
% H_{2^{k-1}} & H_{2^{k-1}} \\
% H_{2^{k-1}} & -H_{2^{k-1}}
% \end{pmatrix} 
% $ and $H_1 {=} \begin{pmatrix} 1 \end{pmatrix}$.
% Also, $(\frac{1}{\sqrt d}H) \cdot (\frac{1}{\sqrt d}H)^T {=} I$ and $\mathit{det}(\frac{1}{\sqrt d}H) \in [-1,1]$.} 
\end{definition}

\begin{definition}[\cite{10.1145/1132516.1132597}]
A randomized Hadamard transform, $\mathcal{H}$, of a vector, $x \in \mathbb{R}^{2^k}$, is defined as $\mathcal{H}(x) \coloneqq H_{2^k}Dx$, where $H_{2^k}$ is a normazlized Walsh-Hadmard matrix, and $D$ is a diagonal matrix whose diagonal entries are i.i.d. Rademacher random variables (i.e., taken uniformly from $\{+1,-1\}$). While $\mathcal{H}$ is randomized and thus defines a distribution, when $D$ is known, we abuse the notation and define the inverse Hadamard transform as \mbox{$\mathcal{H}^{-1}(x) \coloneqq (H_{2^k}D)^{-1}x = DH_{2^k}x$}.

% Let $R_H$ denote the rotation matrix $\frac{HD}{\sqrt d}\in\mathbb R^{d\times d}$, where $H$ is a Walsh-Hadamard matrix and $D$ is a diagonal matrix whose diagonal entries are i.i.d. Rademacher random variables (i.e., taking values uniformly in $\pm 1$). 
% Then $\mathcal R_H(x)=R_H \cdot x =  \frac{1}{\sqrt d} H \cdot (x_1 \cdot D_{11}, \dots, x_d \cdot D_{dd})^T$ is the randomized Hadamard transform of $x$ and $\mathcal R_H^{-1}(x)=R^T_H \cdot x = \frac{DH}{\sqrt d} \cdot x$ is the inverse transform.%\SV{maybe define the inverse one as well?}
\end{definition}

The quantization operates as follows.
Given a vector, denoted $x\in \mathbb{R}^d$, we first \emph{precondition} it using a randomized Hadamard transform, $\mathcal{H}$, and normalize by multiplying by $\nicefrac{\sqrt{d}}{\norm{x}_2}$. There are several desired outcomes of this transform\footnote{We also note that the transform has the advantage of having a vectorized, in-place, $O(d\log d)$-time implementation~\cite{fino1976unified}.}. First, the dynamic range of the values is reduced (measured, for instance, by the ratio of the $\ell_\infty$ and the $\ell_2$ norms). Loosely speaking, we can think of the transform as spreading the vector's information evenly among its coordinates. Second, regardless of the distribution of the input vector, each coordinate of the transformed vector will have a distribution that is close to the standard Gaussian distribution (as an outcome of the central limit theorem).
%\amit{ from 'reduces the error incurred by subsequent coordinate-wise qu antization' to something else perhaps in the lines of "and pushes $\norm{x}_\infty$ to its lower bound of $\norm{x}_2/\sqrt{d}$, which benefits subsequant quantization"}. 
%\amir{@amit - how about the following approach: we transform the vector $x$, and normalize it by $\norm{x}_2/ \sqrt{d}$, call it $y$. Then, $y$ will look close to $\mathcal{N}(0, 1)$. Therefore we can design a scalar quatizer that is matched to this distribution. This also simplifies the algorithm (IMO): step 1: y = H(X), step 2: encode $y$ (and the argmax equation is simpler), step 3: send $X,\norm{x}$.}
After the transform, we perform scalar quantization that is optimized for the $\mathcal{N}(0,1)$ distribution, using $K$-means (also known as Max-Lloyd in the quantization literature \cite{gershogray92}), with $K=2^B$. 
%scalardeterministically assign coordinates to integers in $[0, \dots, 2^B-1]$ \amit{@amir CLT comment replacing "distribution is similar in expectation to the normal distribution $\mathcal{N}(0, \norm{x}_2/ \sqrt{d}$)"} by assigning each coordinate to the nearest \mbox{$K$-Means} cluster, where the number of clusters is $2^B$, their centroids are precomputed on the normal distribution, and then scaled by $\norm{x}_2 / \sqrt{d}$. 
The vector $X$ of cluster assignments together with the original vector's $\ell_2$ norm can now be stored as the compressed representation of the original vector.

% \noindent \textbf{Deterministic K-Means assignment.}
 %recent method called DRIVE~\cite{vargaftik2021drive}. In their paper, the authors analyse the Hadamard transform and argue
% After preconditioning, we deterministically assign coordinates to integers in $[0, \dots, 2^B-1]$ 
% % following an discussion by the authors of the DRIVE paper ~\cite[Section 6]{vargaftik2021drive}, 
% \amit{todo CLT comment}
% who argue that it is reasonable to assume that the post-Hadamard coordinate distribution is similar in expectation to the normal distribution $\mathcal{N}(0, \norm{x}_2/ \sqrt{d}$). Given this observation, they suggest assigning each coordinate to the nearest \mbox{$K$-Means} cluster, where the number of clusters is $2^B$, their centroids are precomputed on the normal distribution, and then scaled by $\norm{x}_2 / \sqrt{d}$. The vector of cluster assignments together with the original vector's l2 norm can now be stored as the compressed representation of the original vector. 

To retrieve an estimate of the original vector, we perform the same steps in reverse. We replace the vector of cluster assignments $X$ with a vector $\hat y$ containing each assigned cluster's centroid, denormalize, and then apply the inverse randomized Hadamard transform, $\mathcal{H}^{-1}$. To avoid encoding $D$ directly, we recreate it using  \emph{shared randomness} \cite{newman1991private} (e.g., a shared pseudorandom number generator seeded from a hash of the vector's text).
Different variations of the quantization scheme are discussed in \S~\ref{subsec:quantization-evaluation}. 
% \alex{what would happen if you hadn't applied the hadamard transform?}

\paragraph{\textbf{Block-wise Quantization.}} The AESI encoder reduces the dimension of the contextual embeddings from hundreds (e.g., 384) to a much smaller number (e.g., 12). On the other hand, the randomized Hadamard transform's preconditioning effect works best in higher dimensions \cite{10.1145/1132516.1132597}. In order to resolve this conflict, we first concatenate the reduced-dimension vectors of all the tokens from a single document. We then apply the Hadamard transform with a larger block size (e.g., 128) on the concatenated vector, block-by-block (padding the last block with zeros when necessary). When evaluating the compression efficiency, we consider the overhead incurred from (a) the need to store the vectors' $\ell_2$ norms and (b) the padding of the final Hadamard block in a concatenated vector. Balancing these factors should be done per use case.

\begin{algorithm}[t]
\caption{~$B$-bits Vector Quantization (DRIVE) \cite{vargaftik2021drive}}
\begin{flushleft}
$\mathcal H$ - A randomized Hadamard transform \\
$\text{\LARGE $\mathfrak{c}$}$ - $K$-Means centroids over the normal distribution, where $K=2^B$
\end{flushleft}
% \begin{multicols}{2}
\begin{algorithmic}[1]
\vspace{1mm}
  \Statex \hspace*{-4mm}\textbf{Quantize($x \in \mathbb{R}^d$):}
  \vspace{0.5mm}
  %\vspace*{1mm}
  \Statex $y:= \frac{\sqrt{d}}{\norm{x}_2}\mathcal H(x)$
  %\vspace*{1mm}
  \Statex Compute $X_i = \argmin_k \left| y_i  -  \text{\LARGE $\mathfrak{c}$}_k\right|$: \mbox{$X\in \{0,\dots,2^B-1\}^d$}
% \begin{align*}

% \end{align*}
%   \vspace*{-1.5mm}
   \Statex \Return $X$, $\norm{x}_2$
\end{algorithmic}
\vspace{2mm}
\begin{algorithmic}[1]
  \Statex \hspace*{-4mm}\textbf{Dequantize($X$, $\norm{x}_2$):}
  \vspace{0.5mm}
  \Statex Compute $\hat y_i =   \text{\LARGE $\mathfrak{c}$}_{X_i}$: \mbox{$\hat y \in \{\text{\LARGE $\mathfrak{c}$}_0,\dots,\text{\LARGE $\mathfrak{c}$}_{2^B-1}\}^d$}
% \begin{align*}
% \hspace*{-4mm}
% \hat y_i =   \text{\LARGE $\mathfrak{c}$}_{X_i}
% \end{align*}
% \vspace*{-3.5mm}
   \Statex \Return $ \hat x = \mathcal H^{-1}\left(\frac{\norm{x}_2}{\sqrt{d}} \hat y\right)$
\end{algorithmic}
\label{code:ourQ}
\end{algorithm}

\section{Experimental Settings} \label{subsec:exprsetup}

In this section, we describe the tasks and datasets used to evaluate the competing approaches, which are evaluated in terms of their quality to rank high relevant text passages for a given query. Next, we describe the baseline and the different configurations of SDR with emphasis on how we measure the compression ratio.

\subsection{Evaluation Metrics}

We consider two groups of evaluation metrics that capture different aspects of the resulting embeddings, namely:

\textbf{Ranking:} To measure the quality of the reconstructed token embeddings from their compressed vectors $c$, we consider the two official evaluation metrics in MSMARCO~\cite{nguyen2016ms}: MRR@10 and nDCG@10 (further discussed below).

\textbf{Compression:} We measure \emph{Compression Ratio} as amount of storage required to store the token embeddings when compared to the 
distilled baseline. %, which represent each token as a 384-dim vector. 
%naive baseline, which in this case is storing the full 768-dim token embedding from BERT. 
%E.g., $CR=2$ translates that we require twice as less as the baseline to store the token embeddings.
E.g., $CR=10$ implies storage size that is one tenth of the baseline vectors.

\subsection{Tasks and Datasets} 
\label{subsec:tasks}
To evaluate the effectiveness of our approach SDR and the competing baseline, we use the passage reranking task of MSMARCO~\cite{nguyen2016ms}. 
In this task, we are given a query and a list of 1000 passages (retrieved via BM25), and the task is to rerank the passages according to their relevance to the query. 
We consider two query sets:

\T{MSMARCO-DEV} We consider the development set for MSMARCO passage reranking task, which consists of 6980 queries. 
On average, each query has a single relevant passage, and other passages are not annotated. 
The performance of models is measured using the mean reciprocal rank, MRR@10, metric. \\
\T{TREC 2019 DL Track} Here we consider the test queries from TREC 2019 DL Track passage reranking dataset. 
Unlike the above, there are many passages annotated for each query, and there are graded relevance labels (instead of binary labels). This allows us to use the more informative nDCG@10 metric. 
Due to the excessive annotation overhead, this dataset consists of just 200 queries, so results are more noisier compared to MSMARCO-DEV. 
% @besnik: this description is incorrect. 
% To evaluate the effectiveness of our proposed compression approach SDR and our competing baseline, we consider two tasks from the MSMARCO 2019 Deep Learning Track~\cite{nguyen2016ms}: 

% \begin{enumerate}[label=\textbf{Task\#\arabic*:},leftmargin=*]
%     \item Here, we consider document reranking task, where for a given list of top-1000 documents (retrieved via BM25), the task is to rerank the documents according to their relevance to a given query. The performance of models is measured using the mean reciprocal rank, MRR@10, metric. This task assess the suitability of the competing approaches in capturing the document context for a given query. We use the development set for this task, which consists of 6980 queries and the corresponding top--1000 documents.
%     \item Similar as above, here we consider the passage re-ranking task, where for a given top--1000 passages the task is to rerank them such that most relevant passages for a query are ranked highest. This performance of models is measured using the normalized discounted cumulative gain, $nDCG@10$, metric, and assess whether the models can capture the graded passage relevance for a given query. For this task we use the test subset consisting of 200 queries and the corresponding top--1000 passages.
% \end{enumerate}

\subsection{Baseline -- \emph{\splitbert}} \label{subsection:baseline}
Our algorithm is based on the late-interaction architecture \cite{macavaney2020efficient,gao2020earl,nie2020dc,chen2020dipair,cao2020deformer} depicted in Figure~\ref{fig:late-interaction}. 
We created a model based on this architecture, which we name \splitbert, consisting of 10 layers that are computed independently for the query and the document with an additional two late-interaction layers that are executed jointly. We initialized the model from pre-trained weights\footnote{https://huggingface.co/cross-encoder/ms-marco-MiniLM-L-12-v2} and fine-tuned it using knowledge distillation from an ensemble of BERT-Large, BERT-Base, and ALBERT-Large \cite{hofstatter2020interpretable} on the MSMARCO small training dataset, which consists of almost 40M tuples of query, a relevant document, and an irrelevant document.\footnote{The checkpoint will be released with the published paper.}

% \WIP{ 
% \nachshon{low priority: Consider https://huggingface.co/cross-encoder/ms-marco-MiniLM-L-2-v2 model. }
% }

\subsection{SDR Configuration and Training}
We denote the SDR variants as ``\mbox{\oursabbr-\{c\}-\{B\}b}'' where \{c\} is replaced with the width of the encoded vector and \{B\} is replaced with the number of bits in the quantization scheme. When discussing \oursabbr with no quantization, we simply write ``\mbox{\oursabbr-\{c\}}''.

%\paragraph{\textbf{Training.}} SDR requires a pre-training of its autoencoder. This is done in order to minimize the reconstruction error of the encoded context vector $c$ during the decoding phase.
\paragraph{\textbf{Training.}} SDR requires a pre-training of its autoencoder to minimize its reconstruction error.
We train the autoencoder on a random subset of 500k documents from the total of 8.8M documents present in MSMARCO collection to reduce training time. 
%This in turn reduces training time, and as we will see in \S\ref{sec:eval} has no quality implications.
%\nachshon{We cannot say this without showing it has not quality implication. Ok to omit?}

\paragraph{\textbf{Quantization overhead.}} We incorporate the quantization overhead into the computation of the \emph{compression ratios} as follows: (a) we assume that the additional DRIVE scalar per block (the \mbox{$\ell_2$-norm}) is a float32 and get a space overhead of $32/(\left<\operatorname{block-size}\right>\cdot B)$ {where $B$ is the number of bits in the quantization scheme}; and (b) we consider the overhead caused by padding, which depends on the length distribution of documents in the dataset. 
For the sake of simplicity, we fixed the block size at 128 and measured the padding overhead using a random sample of 100k documents from the MSMARCO-DEV dataset. This measured padding overhead is
20.1\%,  9.7\%, 6.7\%, and 4.5\%, for \oursabbr 4, 8, 12, and 16, respectively. 
In \S~\ref{subsec:quantization-evaluation}, we discuss possible means to reduce this overhead.

\section{Evaluation}\label{sec:eval}

In this section, we present the main results on compression ratios and quality tradeoff of the SDR scheme (\S~\ref{subsec:main-evaluation}). Later, we examine how the proposed autoencoder (\S~\ref{subsec:auto-encoder-mrr}) and quantization (\S~\ref{subsec:quantization-evaluation}) compare to other baselines. 
Finally, in \S~\ref{subsec:analysis-vectors}, we provide insights and discuss the information captured by our \oursabbr{}-encoded vectors.

% Building on the evaluation of our method, in this section we aim to provide insights on complementary topics. We start by demonstrating the necessity for this work by evaluating how the document embedding size affects fetching latency (\S~\ref{subsec:analysis-elasticsearch}). Then in \S~\ref{subsec:analysis-vectors}, we discuss on the information captured by our \oursabbr{}-encoded vectors.

\subsection{Main results} \label{subsec:main-evaluation} % \subsection{Experimental Settings} \label{subsec:evaluation-settings} 

\begin{table}[]
\setlength{\tabcolsep}{0.3em}
\resizebox{1\columnwidth}{!}{
	\begin{tabular}{c|c|c|l|l}
	\toprule
		\makecell{Quant. \\ bits ($B$)} & \makecell{\oursabbr \\ dim. ($c$)}      &  \makecell{Comp. \\ ratio (CR)} & \makecell{\textsc{msmarco-dev} \\ MRR@10}  & \makecell{TREC19-DL \\ nDCG@10}   \\
		\midrule
		\midrule
%  		& \splitbert      & 2 & 0.3768                   & \\
 		\multirow{4}{*}{\makecell{32\\ (float)}}  
% 		0.20\%, 0.0016\%$'$  
		& 16           & 24 & 0.3759 (-$0.0009$\ignore{, $-0.0042\nicefrac{\%}{\text{CR}}$})     &  0.772 (-$0.002$)            \\
		& 12         & 32 & 0.3725 (-$0.0043$\ignore{, $-0.0178\nicefrac{\%}{\text{CR}}$})$^{*}$       & 0.784 (+$0.01$)         \\
		& 8          & 48 & 0.3711 (-$0.0057$\ignore{, $-0.0157\nicefrac{\%}{\text{CR}}$})$^{*}$ & 0.781 (+$0.007$)   \\
		& 4         & 96 & 0.3660 (-$0.0108$\ignore{, $-0.0149\nicefrac{\%}{\text{CR}}$})$^{*}$        & 0.775 (+$0.001$) \\
		\hline
		\multirow{4}{*}{6}
		& 16 &  121 &  0.3753 (-$0.0015$\ignore{, $-0.0016\nicefrac{\%}{\text{CR}}$})  & 0.772 (-$0.002$) \\
		& 12 & 159 & 0.3728 (-$0.004$\ignore{, $-0.0034\nicefrac{\%}{\text{CR}}$})$^{*}$      & 0.780 (+$0.006$)  \\
		& 8  & 231 & 0.3689 (-$0.0079$\ignore{, $-0.0045\nicefrac{\%}{\text{CR}}$})$^{*}$    & 0.775  (+$0.001$)   \\
		& 4 & 423 & 0.3624 (-$0.0144$\ignore{, $-0.0045\nicefrac{\%}{\text{CR}}$})$^{*}$     & 0.766 (-$0.008$)   \\
		\hline
		\multirow{4}{*}{5}
		& 16 & 145 & 0.3735 (-$0.0033$\ignore{, $-0.0030\nicefrac{\%}{\text{CR}}$})$^{*}$           & 0.772  (-$0.002$)         \\
		& 12 & 190 & 0.3714 (-$0.0054$\ignore{, $-0.0038\nicefrac{\%}{\text{CR}}$})$^{*}$        & 0.778  (+$0.004$)          \\
		& 8 & 277 & 0.3649 (-$0.0119$\ignore{, $-0.0057\nicefrac{\%}{\text{CR}}$})$^{*}$       & 0.770  (-$0.004$)  \\
		& 4 &  506 &  0.3540 (-$0.0228$\ignore{,   $-0.0060\nicefrac{\%}{\text{CR}}$})$^{*}$       & 0.767  (-$0.007$) \\
		\hline
        \multirow{4}{*}{4}
        & 16  & 181 & 0.3665 (-$0.0103$\ignore{, $-0.0076\nicefrac{\%}{\text{CR}}$})$^{*}$       & 0.766 (-$0.008$) \\
        & 12   & 236 & 0.3639 (-$0.0129$\ignore{, $-0.0072\nicefrac{\%}{\text{CR}}$})$^{*}$      & 0.764  (-$0.01$)    \\   
        & 8 & 344 & 0.3544 (-$0.0224$\ignore{, $-0.0086\nicefrac{\%}{\text{CR}}$})$^{*}$           & 0.765 (-$0.009$)    \\ 	
        & 4  & 629 & 0.3408 (-$0.036$\ignore{, $-0.0076\nicefrac{\%}{\text{CR}}$})$^{*}$         & 0.752 (-$0.022$)$^{*}$ \\  
        \hline \hline
        \multicolumn{2}{c|}{\splitbert (Baseline)} & 1 & 0.3768 & 0.774  \\
		\bottomrule
	\end{tabular}}
	\caption{\emph{SDR performance in various configurations}: MRR@10 and nDCG@10 are measured over MSMARCO, as described in \S~\ref{subsec:tasks}. The absolute difference w.r.t. the \splitbert baseline is shown in parentheses. We measured statistical significance using relative paired t-test\protect\footnotemark{} and denote with $^*$ cases with $p < 0.05$. Note that \oursabbr{}-16-6b shows no significant drop in ranking quality. The compression ratios indicate the reduction in storage size, including padding and normalization overheads.}
	\label{table:main-results}
\end{table}
	
	%Using paired t-test statistic we measure statistical significance and denote with $^{*}$ significant ($p < 0.05$) the results with means different than the baseline. 
	%In terms of compression rate reductions, we measured the reduction in contrast to storing the full 768-dimensions from BERT using \texttt{float32} and considering padding and normalization overheads. 
	%\WIP{\nachshon{low priority: consider t-test with margin.}
% 	The highlighted cells are examples of efficient configurations, the first provides 243x compression without a significant lost, and the second provide 1013x compression with only 6.05\% degradation
Table~\ref{table:main-results} shows the results on both query sets for SDR and its compression ratio against storing contextual token embeddings uncompressed. In terms of compression ratio, it can be seen that \oursabbr{} allows us to massively reduce storage requirements both with and without quantization. 

\textbf{SDR without Quantization:} \oursabbr-16 reduces storage requirements by 24x with an insignificant performance drop in MRR@10 and nDCG@10 compared to the baseline. Higher compression rates can be achieved by further reducing the \emph{context vector}'s dimensions ($c \in \{4, 8, 12\}$) at the cost of having a small but statistically significant lower performance than the baseline. Depending on the use case, such tradeoffs are highly desirable, allowing for extreme compression rates that minimize the costs of deploying Q\&A systems. 
For instance, \oursabbr-12 achieves a 32x compression rate, and although the performance drop is statistically significant, the absolute difference is just 0.0043 for MRR@10 (with a tiny increase for nDCG@10). In some situations, a better compression rate would justify this slight reduction in performance.
%For instance, \oursabbr-12 achieves a 32x compression rate, and although it has significantly lower performance on both tasks, in terms of absolute differences with 1.14\% and 3.5\% for MRR@10 and nDCG@10, for hundreds of queries and reranking of 1000 documents/passages, this presents a satisfactory performance, that most likely will not have any observable impact in addressing user's query needs. 
Lastly, when using only 4 dimensions for the context vector, we obtain nearly 100x compression rate, fitting the entire MSMARCO collection in less than 11GB of memory.

\footnotetext{A relative t-test compares systems on the same queries, making it easier to distinguish subtle differences. An independent t-test corresponds to an external observer that only sees results on one system at a time. According to the latter, most of the cases are indistinguishable, except for AESI-\{4, 8\}-4b and AESI-4-5b.}

\textbf{SDR with Quantization:} Highly efficient compression rates and reranking performance is achieved when using quantization techniques. For instance, \oursabbr-16-6b reaches a compression rate of 121x, including padding and normalization overheads, while at the same time showing no significant ranking performance drop.  Using \oursabbr-16-6b, a document's embedding can be stored with only 947 bytes, which, as shown in Table~\ref{table:elasticsearch-latency} in Appendix~\ref{subsec:analysis-elasticsearch}, does not add a significant network latency cost in fetching such vectors. The entire MSMARCO collection can be stored within 8.6GB. There are several advantages of fitting the entire collection's representation into the main memory of the hosting machine, allowing for fast access, further fine-tuning, etc. If further compression rates are required, \oursabbr-8-5b uses just 5 bytes per token, reaching a compression rate of 277x and 487 bytes per document on average. At this level of compression, the entire MSMARCO corpus fits in 3.8GB.
The MRR@10 drop is noticeable (0.0119) but still quite low. 
Finally, for TREC19-DL, the impact of compressing token embeddings is less evident. Only in the most extreme cases such as \oursabbr-4-4b we see a significant drop in nDCG@10 performance. %These results are very encouraging given that in this task the models need to capture the graded passage relevance w.r.t. a given query. 
These results demonstrate that the performance drop is very small, showing the effectiveness of our method.

% DRIVE-B is also an effective strategy for dropping storage requirements, but does not show similar effectiveness to the auto-encoder results. DRIVE-4 provides 8x reduction in storage at \nachshon{Complete once we have numbers. }
%DRIVE-B is also an effective strategy for dropping storage requirements. DRIVE-8 show no loss in MRR. DRIVE-4 reduces storage requirements to 4 bits per float (8x reduction) with very minimal loss in MRR, which is not statistically significant. 
%Even DRIVE-2 drops quality only minimally, but the drop is noticable. 
% \nachshon{Consider returning DRIVE back, then keeping the commented out explanation. Otherwise, we need to rephrase next paragraph.}
% compared to naively storing BERT 
% (i.e., 12 floats with 4 bits per float, or equivalent to 1.5 floats). 

\subsection{Autoencoder Evaluation} \label{subsec:auto-encoder-mrr}
To better understand the impact of the autoencoder, we present MRR@10 results as a function of autoencoder dimensions (i.e., number of floats stored per token) and with the different autoencoder configurations. In addition to the 2-layer \oursabbr{} architecture we described in \S~\ref{subsec:autoencoder} (\textbf{\oursabbr{}-2L}), we consider the following variations:
% \nachshon{@amit to modify format to be similar to 4.4.}
\\\\
\T{AutoEncoder with 2 Layers (AE-2L).} Standard 2-layer autoencoder with $\operatorname{gelu}$ activation. This is the same as AESI, only without the side information. %\besnikf{What does links mean here? \amit{the "skip-connections" from level-0 (the arrows going out from the 'side information' box in Figure 3}}
	\ignore{\oren{did u tie the weights between the encoder and decoder?
Did u use a normalization layer in the bottleneck in any of the baselines? }}

\T{AutoEncoder with 1 Layer (AE-1L).} Standard autoencoder with a single dense layer in the encoder and decoder.

\T{AESI with 1 Layer (AESI-1L).} \oursabbr{} with a single dense encoder and decoder layer with side information.
%\nachshon{I removed the work link since it does not help understanding. }

\T{DECoder-only AESI (AESI-DEC-2L).} Similar to \oursabbr{}, however the encoder has no access to the side information. Recall that the static token embeddings are required by the decoder to help reconstructing the original vector. This variant checks if the static token embeddings help in the encoding part
\\
%The uncontextualized token embeddings is mostly important for the decoder, so that the original vector can be reconstructed without relying just on the (few dimension) encoded vector. 
%Therefore, we experiment with an autoencoder where the uncontextualized token embeddings is only provided to the decoder. 
%We further experiment with a single-layer autoencoder with uncontextualized token embeddings, where the 2 layers with a non-linear activation is replaced with a single layer.
%This provides a minor improvement in running time, but is unable to model non-linear combination of the compressed vector with the uncontextualized token embeddings. 

\begin{figure}
    \centering
    \includegraphics[width=1\columnwidth]{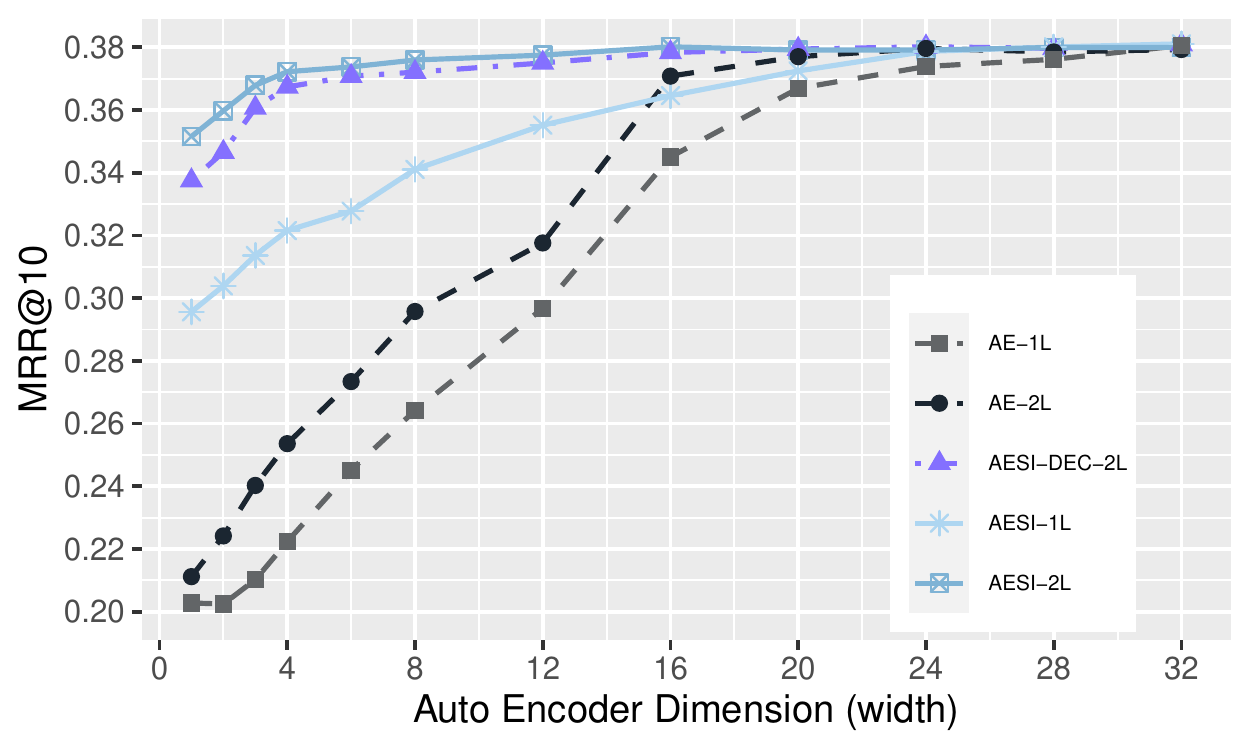}
    \caption{MRR@10 was measured on the MSMARCO-DEV-25 dataset as a function of autoencoder dimensions. The results are shown for standard autoencoders (AE) and our approach (AESI), with single or two-layer encoder and decoder networks. The x--axis shows the dimension of the encoded vector $c$. }
    \label{fig:auto-encoder-mrr}
\end{figure}

To reduce measurement overhead, we ran the experiment over the top 25 \splitbert{} passages for each query, denoted \textsc{msmarco-dev-25}, which has a negligible impact on the results. 
Figure~\ref{fig:auto-encoder-mrr} shows the results for the different autoencoder configurations. 
Providing the side information to the autoencoder proves to be very effective in reducing storage costs, especially when the encoded vector size is small. 
%Providing the side information to the autoencoder proves to be very effective in reducing storage costs. 
%Furthermore, for a small number of dimensions ($\leq 16$), the side information allows us to obtain better MRR@10, when compared to the standard 2-layer autoencoder. 
A 2-layer encoder/decoder model, as expected, is more effective than a single-layer model. The gap is especially large when using side information, showing that the interaction between the encoded vector and the static token embeddings is highly nonlinear. Finally, we note that it is possible to provide static token embeddings only to the decoder, but providing it also to the encoder slightly increases the overall MRR@10 score. 

% It can be seen that the side information we provide to the autoencoder (namely, the static token embeddings) is very effective in reducing storage requirements,  achieving much better MRR@10 when the number of dimensions is small compared to a standard 2-layer autoencoder. 

\subsection{Quantization Evaluation} \label{subsec:quantization-evaluation}\label{sec:altQ}

To study the impact of quantization, we fix \oursabbr{}-16 as our baseline and measure how different quantization strategies and number of bits affect the MRR@10 score. Note that we do not measure quantization over the baseline \splitbert since it can only achieve a compression ratio of up to 32x per coordinate (using 1 bit per coordinate). In addition to {\bf DRIVE} (\S~\ref{subsec:quantization-algorithm}, Algorithm~\ref{code:ourQ}), we consider the following quantization strategies:
\bigskip

\T{Deterministic Rounding (DR) \cite{gershogray92}.}
Maps the input coordinates into the $[0, 2^B-1]$ range using min-max normalization and rounds to the nearest integer.

\T{Stochastic Rounding (SR) \cite{barnes1951electronic, doi:10.1137/20M1334796}.}
Normalizes as before using min-max normalization, and additionally adds a uniform \emph{dither} noise in $(-0.5, 0.5)$ and then rounds to the nearest integer. This provides an unbiased estimate of each coordinate.

\T{Subtractive Dithering (SD) \cite{1057702, 256489}.}
Same as SR, only now before denormalization, instead of just using the values in \mbox{$\{0,\dots,2^B-1\}$}, we first subtract the original dither noise, which we assume can be regenerated using shared randomness. This is an unbiased estimator with reduced variance.

\T{Hadamard Variants (H-DR, H-SR, and H-SD).}
These variants correspond to the previous methods; only they are preceded by a randomized Hadamard transform.

\T{DRIVE with Bias Correction (DRIVE-BC) \cite[Appendix C.3]{vargaftik2021drive}.}
This variant of DRIVE optimizes for lower bias over the mean squared error (MSE) by multiplying the dequantization result in Algorithm \ref{code:ourQ} by a bias correction scalar: $\norm{x}^2_2/\norm{\hat{y}}^2_2$ .

\bigskip

\begin{figure}
    \centering
    \includegraphics[width=3.2in]{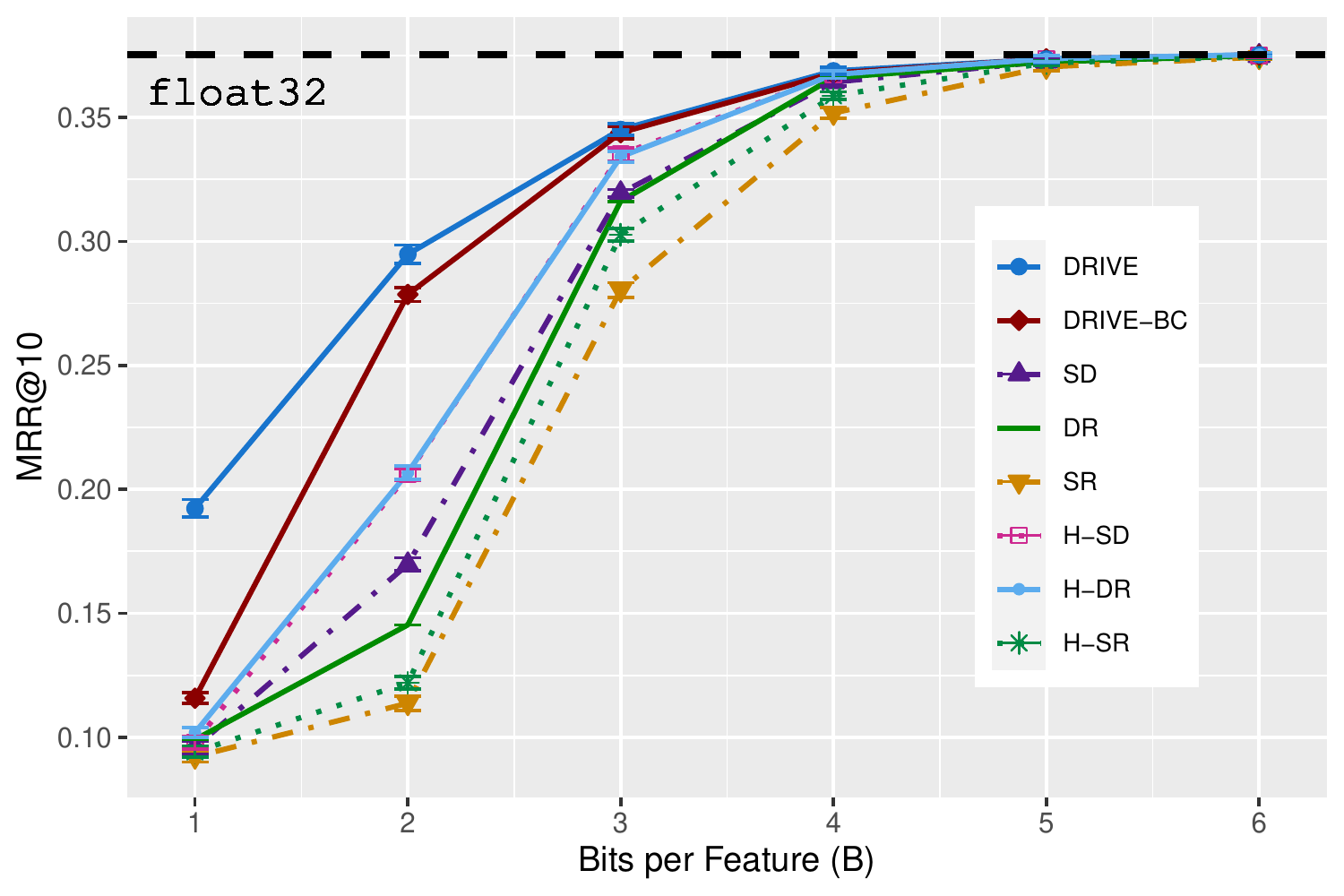}\vspace{-10pt}
    \caption{MRR@10 for different quantization methods. Each run quantizes and dequantizes \oursabbr{}-16 encoded documents over the MSMARCO-DEV-25 dataset. For each randomized quantization method and number of bits, we take the average of 10 runs (the error bars show the standard deviation). 
    % \amit{@besnik create with a similar style, but match the existing content, i.e., sixth bit, error-bars, rename DRIVE->DRIVE-BC, rename DRIVE-NBC->DRIVE, remove H\_SR2}
    % \amit{@besnik larger fonts? (legend/axes labels/float32 line)}
    }
    \label{fig:compress_algo_analysis}
\end{figure}

Figure~\ref{fig:compress_algo_analysis} shows the results for the different quantization methods. First, we observe that the Hadamard variants perform better than their non-Hadamard counterparts. Second, we see that DRIVE performs better than all other schemes. The differences are more pronounced in the low-bit regime, where the choice of quantization scheme has a drastic impact on quality\ignore{; when the number of bits is large (e.g., 6), all methods perform reasonably well, and the gap between the quantization schemes is reduced}. We also note that unlike in other use cases, such as distributed mean estimation, bias correction is inappropriate here and should not be performed at the cost of increased mean squared error (MSE). This conclusion follows by observing that DRIVE and the deterministic rounding methods (DR, H-DR) are respectively better than DRIVE-BC and the stochastic rounding methods (SR, H-SR). We add that the subtractive dithering methods (SD, H-SD), expectedly, work the same or better than their deterministic counterparts since they produce a similar MSE while also being unbiased.

The current quantization scheme requires padding to full 128 blocks. For 4-8 \oursabbr{} size, the padding overhead may reach 10\% -- 20\% percent. In addition, we send a normalization value per 128-block, which we currently send as a float32 value, adding 4\% -- 5\% additional overhead. Padding can be reduced by treating the last 128-block separately, e.g., applying a method that does not require Hadamard transform. Normalization overhead can be reduced, e.g., by sending normalization factors as float16 instead of full float32.  However, such solutions complicate the implementation while providing limited storage benefits, hence, they were not explored in the context of this paper. 

\textbf{Beyond Scalar Quantization.}
Scalar quantization using a fixed number of bits is a suboptimal technique in general since it does not allocate fewer bits for more frequent cases. 
Entropy coding\cite{gershogray92} can do better in this aspect. However, for the 6-bits case, the measured entropy was just 5.71 bits, so the potential improvement from variable-length coding does not seem to justify the efforts (even before accounting for the overhead incurred by Huffman coding or arithmetic coding). 
Another direction is to use vector quantization or entropy-constrained vector quantization. To estimate an upper bound on the benefits of such techniques, we use the rate-distortion theory (from the information theory field), which studies the optimal tradeoffs between distortion and compression rate \cite{cover_thomas}. 
For a Gaussian source, which is a reasonable approximation of the output of Hadamard transform, it is known that the optimal (lossy) compression rate is given by $\frac{1}{2}\log_2(\frac{1}{MSE})$. 
Compared to the 6-bits case, the optimal rate is 5.35 bits, indicating an upper bound of 11\%; results for different bit rates are similar. 
Therefore, we conclude that the limited gains do not seem to justify the added system complexity.

\subsection{Intrinsic Evaluation of \oursabbr{}-Encoded Vectors}\label{subsec:analysis-vectors}

In the previous sections, we showed the effectiveness in ranking and utility in compression rates of \oursabbr{} over AE architectures. 
However, such evaluations do not capture the encoded information at the token-level. In this intrinsic evaluation we try to discern when and why adding the static embedding as side information contributes to better capturing the token meaning.

We study the effectiveness of different autoencoder configurations in reconstructing back the original token vector, as measured through the MSE between the original vector and the reconstructed vector:
\begin{equation*}
MSE\left(v, D\left(E(v, u), u\right)\right)\ ,
\end{equation*}
where $v$ is a contextualized vector (\splitbert{} output at layer $10$), $u$ is the static embedding, and the encoder $E(v, u)$ and the decoder $D(e, u)$ are as defined in \S~\ref{subsec:autoencoder}. High MSE scores indicate the inability of the autoencoder to encode the original vector's information.

\textbf{Document Frequency:} One way to assess the importance of a document w.r.t. a query is through the inverse document frequency of query tokens, typically measured through TF-IDF or BM25 schemes~\cite{robertson2009probabilistic}. In principle, the more infrequent a query token is in a document collection\shorten{(i.e., high IDF)}, the higher the ranking of a document containing that token will be. Tokens with (very) high frequencies are typically stop words or punctuation symbols, which have lower importance when determining the query-document relevance. %OK to omit (i.e.., high IDF) if space is needed. However, just (IDF) is insufficient.

\begin{figure}
	\centering
	\includegraphics[width=1.0\columnwidth]{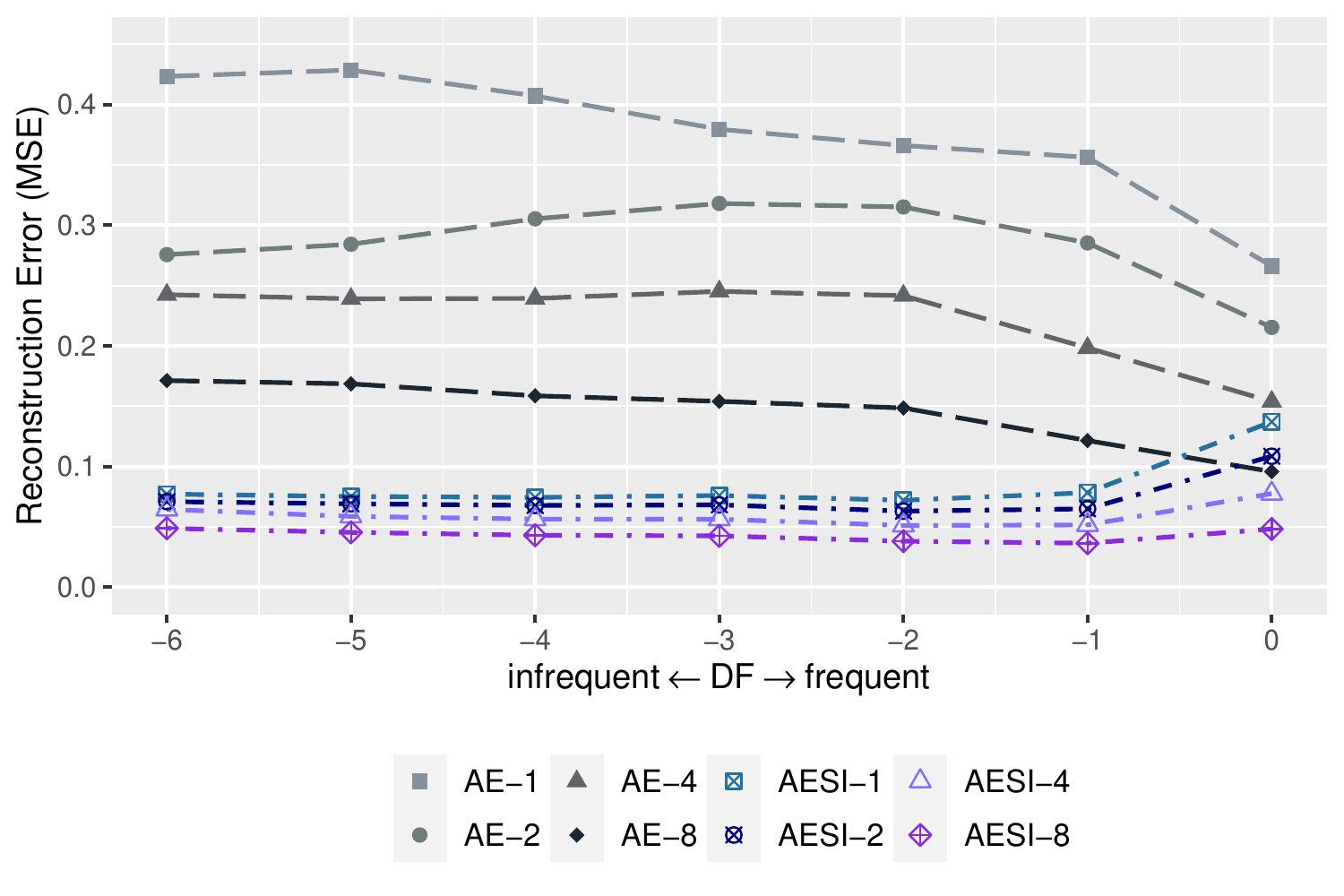}\vspace{-10pt}
	\caption{Reconstruction Error vs. DF for the different \mbox{AE} and \mbox{AESI} configurations. AESI shows robust performance in recovering back the token's representation with a MSE score (y-axis), which is constant for documents with varying DF scores. It is interesting to note that for frequent tokens (i.e., tokens that are function words, hence play a marginal role in retrieval), the error rate is higher when compared to the rest of the tokens. %The line is smoothed using an averaging window of 1 DF.
	}
	
	\label{fig:analysis-idf}
\end{figure}

Based on this premise, we study how MSE varies across token frequency. 
We selected a random sample of 256k documents from MSMARCO, tokenized them, and run them through \splitbert{} to get 20M contextualized token representations. 
Then, for each token we measured their document frequency as $DF(t) = \operatorname{log}_{10}(|\{d \in D:t \in d\}|/|D|)$ (where D is our document collection), and in Figure~\ref{fig:analysis-idf} we plot the average MSE against the rounded DF scores. 
%\footnote{Different colors and shapes indicate the different autoencoder configurations}
From this experiment, we make the following observations. 

First, on all encoded width configurations, our approach, \oursabbr{}, consistently achieves lower MSE compared to the AE architecture (for all DF values). Lower MSE correlates to a better ranking quality, as shown in \S~\ref{subsec:auto-encoder-mrr}. Furthermore, for tokens with low DF, adding the static side information during the training of \oursabbr{} for compression provides a huge advantage, which shrinks when the token is present in many documents in the collection.

Second, on the end spectrum of high-frequency tokens, we note a downwards trend for AE and an upwards trend for \oursabbr{}, especially for $DF \in [-1,0]$.
The MSE decrease for AE is expected since the training data contains more frequent tokens. 
% The MSE increase for \oursabbr{} is intuitive given that in this frequency range, we deal with tokens that are function words and whose role is less tied with the context within the sentence and which reduces the contribution provided by the side information.
%\oursabbr{} is less effective.
% As static embeddings provided as a side information cannot capture context, \oursabbr{} is less effective, hence the slight increase in MSE. 
% \\
% The MSE increase for \oursabbr{} can be explained given that in this frequency range we deal with tokens that are function words and whose role is more in tying up content tokens within a context of the sentence. 
% As static embeddings cannot capture context, this reduces the contribution provided by the side information.
The increase for \oursabbr{} can be explained given that in this frequency range, we deal with tokens that are function words (e.g., `the') whose role is more in tying up content within a sentence and has less standalone meaning. In this case, static embeddings cannot capture context, which reduces the contribution provided by the side information.

\section{Conclusion}
In this paper, we proposed a system called SDR to solve the storage cost and latency overhead of existing late-interaction models. 
The SDR scheme uses a novel autoencoder architecture that uses static token embeddings as side information to improve encoding quality. 
In addition, we explored different quantization techniques and showed that the recently proposed DRIVE (without bias correction) performs well in our use case and presented extensive experimentation. 
Overall, the SDR scheme reduces pre-computed document representation size by 4x--11.6x compared to a baseline that uses existing approaches. 

In future work, we plan to continue investigating means to reduce pre-computed document representation size. % to less than the text size. 
We believe that additional analysis of BERT's vector and their interaction with the context would be fundamental in such an advancement. 

% In this paper, we demonstrated that current late-interaction-based precomputation solutions suffer from increased latency issues that renders them impractical. We solved this issue by providing a novel \ours architecture followed by effective quantization. 

% In future work, we plan to continue investigating means to increase the compression ratio while preserving the model's effectiveness. We believe that further analysis and understanding of the \oursabbr{}-encoded vectors will be fundamental in such an advancement.

%  appendix
\appendix

\section{Appendix: Latency Overhead of Fetching Large Vectors}\label{subsec:analysis-elasticsearch}

\begin{table}[]
    \begin{tabular}{lrr}
        \toprule
        %\begin{tabular}[c]{@{}l@{}}num documents\\ payload size\end{tabular} & \multicolumn{1}{r}{200} & \multicolumn{1}{r}{1000} \\
        payload size & 200 documents & 1000 documents \\
        \hline \hline 
2                                                            & 6.4$\pm$0.8             & 21.9$\pm$1.5             \\
512                                                          & 7.0$\pm$1.1               & 24.9$\pm$0.5             \\
1K                                                           & 7.7$\pm$0.6             & 30.6$\pm$0.5             \\
2K                                                           & 9.7$\pm$0.5             & 42.9$\pm$6.6             \\
4K                                                           & 13.2$\pm$0.6            & 55.1$\pm$1.4             \\
8K                                                           & 21.6$\pm$0.7            & 99.7$\pm$2.8             \\
16K                                                          & 38.4$\pm$1.1            & 191.0$\pm$5.2              \\
32K                                                          & 76.9$\pm$1.9            & 391.8$\pm$11.                   \\
        \bottomrule
    \end{tabular}

% 	\begin{tabular}{lll}
% 	\toprule
% 		blob size (bytes) & time per search query (ms) & std (ms) \\
% 		\hline \hline
% 		2                 & 4.02                      & 0.13    \\
% 		512               & 4.94                      & 0.58    \\
% 		1K                & 4.51                      & 0.27    \\
% 		2K                & 5.03                      & 0.24    \\
% 		4K                & 6.52                      & 0.50    \\
% 		8K                & 9.26                      & 0.47    \\
% 		16K               & 14.78                     & 0.32    \\
% 		32K               & 25.74                     & 0.25   \\
% 	\bottomrule
% 	\end{tabular}
\caption{Elasticsearch retrieval latency (in milliseconds, $\pm$ denotes standard deviation) as a function of payload size and number of fetched documents.}
\label{table:elasticsearch-latency}
\end{table}
% Building on the evaluation of our method, in this section we aim to provide insights on complementary topics. We start by demonstrating the necessity for this work by evaluating how the document embedding size affects fetching latency (\S~\ref{subsec:analysis-elasticsearch}). Then in \S~\ref{subsec:analysis-vectors}, we discuss on the information captured by our \oursabbr{}-encoded vectors.

% \subsection{Latency Overhead of Large Vectors}\label{subsec:analysis-elasticsearch}
Existing work, such as PreTTR \cite{macavaney2020efficient}, argue for (some) compression to reduce storage cost. 
In this appendix, we argue that compression should also be done to reduce the fetching latency of document representation, which, without sufficient compression, can offset any benefits from reducing computation costs. 

Standard ad-hoc retrieval is composed of a retrieval service, such as Elasticsearch\footnote{elasticsearch.com}, which stores the entire corpus, and given an online user, request returns a set of $K$ documents, where $K$ is typically between 100 and 1000. 
Normally, the retrieval service is the only location where the entire corpus is stored, so it is a natural location for storing pre-computed document embeddings. 
Therefore, it is important to understand how retrieval latency changes when the size of pre-computed embeddings varies. 

Towards answering this question, we measured how Elasticsearch retrieval latency varies with different payload sizes.  
The full evaluation setup is deferred to the supplementary material for lack of space.
%omitted for lack of space\footnote{It will appear in the arXiv version of the paper.}. %\nachshon{I think it is important to claim that we will publish the full details.}
The results appear in Table~\ref{table:elasticsearch-latency}.
We note that 1KB roughly corresponds to \oursabbr{}-16-6b, while 512 bytes roughly corresponds to \oursabbr{}-8-6b. 
At this range, the latency increase is minimal. However, the payload size for the baseline system is at least 4x larger, leading to a notable latency increase. 
The original PreTTR work \cite{macavaney2020efficient} considered up to 12x size reduction, leading to 32K document embedding size. 
As can be seen in the table, this results in a prohibitively large latency overhead.

\bibliographystyle{ACM-Reference-Format}
\bibliography{ref}

\section*{Details of The Latency Overhead Evaluation} \label{appendix:latency-details}

First, we set up an Elasticsearch cluster using AWS Elasticsearch service\footnote{https://aws.amazon.com/elasticsearch-service/}. 
This service is used by multiple companies and can be consider as reproducing ``real'' production environment. 
We used default settings for the cluster and set node type to m5.xlarge. 

We then created an Elasticsearch index and populated it with documents. 
To study the effect of fetching latency, we duplicate each document multiple times, each with a different blob (aka payload) size. 
Specifically, we repeatedly select a random sample of 10 words, and denote these as a document. 
For each such document, we consider 8 replications, where for each replica we add a blob field with a string of length X, 
and added a unique size word ``SIZE-X'' to the document, where X is selected from the possible sizes \{2, 512, 1K, 2K, 4K, 8k, 16K, 32K\}. 

Once the index was populated, we run some warm-up queries against the index, to ensure the index is in a steady state. 

During experimentation, we repeatedly query (and fetch) documents with blob size of X and measure how retrieval time changes with X.
Each query is composed of a \texttt{should clause}\footnote{https://www.elastic.co/guide/en/elasticsearch/reference/current/query-dsl-bool-query.html} with 3 randomly selected words, and a \texttt{must clause} with the word ``SIZE-X'', which ensures that only documents with blob size X are returned.
The number of documents to retrieve is limited to 1000 (as the default configuration of MSMARCO passage ranking task) or 200. 
All queries are executed by a single thread to avoid concurrency interference and queuing effects, on an AWS EC2 machine in the same datacenter as the Elasticsearch cluster. 
Each experiment issues 100 queries and is repeated 30 times, while interleaving different X values to avoid potential biases. 
We report the average time in milliseconds (ms) and standard deviation in Table~\ref{table:elasticsearch-latency}. \alex{why table and not a chart?}

\end{document}